\documentclass[11pt,twoside]{article}
\usepackage{./asp2014}

\aspSuppressVolSlug
\resetcounters

\begin{document}

\title{Aggregation and Linking of Observational Metadata in the ADS}
\author{Alberto Accomazzi,
Michael J. Kurtz, 
Edwin A. Henneken, 
Carolyn S. Grant, 
Donna M. Thompson, 
Roman Chyla, 
Alexandra Holachek, 
Jonathan Elliott
}
\affil{Harvard-Smithsonian Center for Astrophysics, 60 Garden Street,
  Cambridge, MA 02138, USA}

\paperauthor{Alberto~Accomazzi}{aaccomazzi@cfa.harvard.edu}{0000-0002-4110-3511}{Harvard-Smithsonian Center for Astrophysics}{ADS}{Cambridge}{MA}{02138}{USA}
\paperauthor{Michael~Kurts}{mkurtz@cfa.harvard.edu}{0000-0002-6949-0090}{Harvard-Smithsonian Center for Astrophysics}{ADS}{Cambridge}{MA}{02138}{USA}
\paperauthor{Edwin~Henneken}{ehenneken@cfa.harvard.edu}{0000-0003-4264-2450}{Harvard-Smithsonian Center for Astrophysics}{ADS}{Cambridge}{MA}{02138}{USA}
\paperauthor{Carolyn~S.~Grant}{cgrant@cfa.harvard.edu}{0000-0003-4424-7366}{Harvard-Smithsonian Center for Astrophysics}{ADS}{Cambridge}{MA}{02138}{USA}
\paperauthor{Donna~M.~Thompson}{dthompson@cfa.harvard.edu}{0000-0001-6870-2365}{Harvard-Smithsonian Center for Astrophysics}{ADS}{Cambridge}{MA}{02138}{USA}
\paperauthor{Roman~Chyla}{rchyla@cfa.harvard.edu}{0000-0003-3041-2092}{Harvard-Smithsonian Center for Astrophysics}{ADS}{Cambridge}{MA}{02138}{USA}
\paperauthor{Alexandra~Holachek}{aholachek@cfa.harvard.edu}{0000-0003-3272-8867}{Harvard-Smithsonian Center for Astrophysics}{ADS}{Cambridge}{MA}{02138}{USA}
\paperauthor{Jonathan~Elliott}{jonathan.elliott@cfa.harvard.edu}{0000-0001-8043-4965}{Harvard-Smithsonian Center for Astrophysics}{ADS}{Cambridge}{MA}{02138}{USA}

\begin{abstract}
We discuss current efforts behind the curation of observing proposals,
archive bibliographies, and data links in the NASA Astrophysics Data
System (ADS).  The primary data in the ADS is the bibliographic
content from scholarly articles in Astronomy and Physics, which ADS
aggregates from publishers, arXiv and conference proceeding sites.
This core bibliographic information is then further enriched by ADS
via the generation of citations and usage data, and through the
aggregation of external resources from astronomy data archives and
libraries.  Important sources of such additional information are the
metadata describing observing proposals and high level data products,
which, once ingested in ADS, become easily discoverable and citeable
by the science community.  Bibliographic studies have
shown that the integration of links between data archives and the ADS
provides greater visibility to data products and increased citations
to the literature associated with them.

\end{abstract}

\section{Introduction}

This paper discusses the current curation of archive bibliographies
and their indexing in ADS.  Integration of these bibliographies
provides convenient cross-linking of resources between ADS and the
data archives, affording greater visibility to both data products and
the literature associated with them.

The primary data curated by ADS is bibliographic information provided
by publishers or harvested by ADS from conference proceeding sites and
repositories.  This core bibliographic information is then further
enriched by ADS via the generation of citations and usage data, and
through the aggregation of external bibliographic information.
Important sources of such additional information are the metadata 
describing high level data products associated with publications,
observing proposals from major missions, the curated
bibliographies for archives and organizations, and the sets
of links between archival observations and published papers. 

While ADS solicits and welcomes the inclusion of this data from US and
foreign data centers alike, the curation of these metadata
is left to the archives which
host the data and which have the expertise and resources to properly
maintain them.  In this regard, the role of ADS is one of resource
aggregation and indexing, providing a lightweight discovery mechanism
through its search capabilities.  While limited in scope, this level
of aggregation can still be quite useful in supporting the discovery
and selection of data products associated with publications.  For
instance, a user can use ADS to find papers which have been classified
in the bibliography for HST, Chandra, and Spitzer, which typically
yields multi-spectral studies making use of data from NASA's Great
Observatories.  Following data links associated with this collection
of papers allows the user to directly download the data products
themselves.

\section{Data Products Indexed in ADS}

In the section below we describe how archives can provide to ADS the
observational metadata which is then integrated in ADS's discovery
platform.  Collecting and sharing this information provides the basis for
the creation of bi-directional links between the ADS and data centers, thus enhancing
the discoverability of data and literature across Astrophysics archives.

\subsection{High-level Data Products}

Important datasets are often described in ``data'' papers, but are also
typically available in machine-readable format as digital surveys or electronic catalogs
hosted by one or more archives (think of the data products generated by
projects such as 2MASS or SDSS).  Additionally, the high-level data
products associated with refereed papers currently being published in
the major astronomy journals are captured and ingested in repositories
such as Vizier, SIMBAD, or NED.  Depending on the nature of the
paper, these datasets might then become available as data collections
which include electronic data catalogs (corresponding to tables or plots in the
paper), or other products such as images and spectra (corresponding to
figures).  

In order to facilitate the discovery of these data products, the ADS
has been indexing these records in its database since 1995, with an
automated feed from Vizier since 2001.  Today, ADS has records for
over 10,000 Vizier catalogs, making this the biggest collection of high-level
data products in our system.  Once they become indexed in ADS, these
records gain greater visibility as well as the ability to be easily
cited in the literature.  It is thanks to this arrangement that
astronomers have been able to cite data catalogs for over two decades,
thus providing a simple (although incomplete) solution to the problem of
data citation.

\subsection{Observing Proposals}

Abstracts for awarded observing proposals are useful additions to the
content indexed in ADS for a variety of reasons.  First and foremost,
they provide early descriptions of current and ongoing science
investigations being carried out by the observing facilities, making
them more discoverable by the research communities.
Observing proposals also provide a direct link to either existing or planned
observations.  Since there can be a significant lag
between the taking of the data and the publication of science
results from its analysis, the observing proposals are often the
only descriptive science metadata associated with the observations
themselves.

Having ADS records for these proposals means that they are not
only easily discoverable but they become part of the scholarly
record, and thus can be formally cited in the literature.  While
this practice is not (yet) mainstream, we note that there are over
350 citations to observing proposals in ADS as of November, 2015.
The ADS currently holds more than 32,000 records for observing
proposals, including more than 9,000 from HST.  The ingest rate has
been above 1,000 per year for more than 15 years.

\subsection{Bibliographies}

The ADS provides a way for users to view curated collections of
records related to a number of well-known astronomy institutions, projects and
repositories.  These collections (commonly referred to in ADS as
``bibliographic groups'') can be used as search filters (``show me all
papers on AGNs which have data from the Hubble Space Telescope'') or
as browsable lists of papers (``show me all papers in the ESO
telescope bibliography'').

The primary goal of such bibliographies has always been to facilitate
the filtering of search results in the ADS, however the
availability of paper-based metrics makes it possible for archivists
and program managers use them in order to assess the impact of an
instrument, facility, funding or observing program.  While this is an
inevitable outcome of the bibliography curation process, we urge
caution whenever impact analysis is performed using bibliographic
data, and encourage users to read the ``fine print'' whenever using and
comparing paper-based metrics. 

It is worth pointing out that while there is now an emerging consensus
within the astronomical library community about the principles behind
these bibliographies \citep{2015ASPC..492...99L}, there is no
single (or simple) set of criteria which is applied across all such
collections.  To promote transparency about the process behind their
creation, the ADS help pages provide a short description of the
bibliographic groups.\footnote{\url{http://doc.adsabs.harvard.edu/abs\_doc/help\_pages/search.html\#Select\_References\_From\_Group}}

\subsection{Data Links}

Links between bibliographic records and data products are a bi-product
of so-called ``telescope bibliographies'' \citep{2012SPIE.8448E..0KA}.  
In creating such a bibliography, the
curator goes beyond the step described above of simply tagging
articles associated with a particular mission or project, and instead
identifies the particular data products analyzed in the paper.  It is
often the case that a paper will use a set of observations from one or
more archives, and therefore an archive's telescope bibliography needs
to record the multiple mappings between a paper and a set of data
products hosted by the archive.  Similarly, an observation or derived
data products may be featured in multiple papers, so the telescope
bibliography needs to support a many-to-many mapping between articles
in ADS and its own datasets.

Given the fact that the list of data products discussed in a paper can
be quite long, the recommended approach for generating these mappings
is having an archive expose all data products associated with a paper
in a single landing page, so that a unique link exists between a
paper and its data products within a given archive.

Exposing the set of curated linkages to ADS is as simple as publishing a table
which correlates a bibliographic identifier (bibcode) with one or more
corresponding resources hosted by the archive, with an optional label
for the anchor text associated with each individual link. 
ADS periodically retrieves and ingests these mappings into its index,
updating the links in an automated fashion, and generating the
corresponding bibliographic groups.

\section{Conclusions}

Being able to read the literature and have access to the data discussed in the paper
be just one click away is obviously of great convenience for the end-user, in addition to
providing evidence in support of the scientific arguments in the paper
and the reproducibility of its results.  However, the integration and 
indexing of this information in ADS allows its use to enhance the discovery
process.  In the introduction we provided an example query based on bibliographic group
indexing which finds multi-wavelength papers appearing in NASA's Great
Observatories bibliographies.  
A few additional examples which use ADS's new search
engine\footnote{\url{https://ui.adsabs.harvard.edu}} 
illustrate the kind of inquiries which become possible once
this information is indexed:
literature-based topic searches constrained to a particular 
collection (\emph{exoplanets and bibgroup:Spitzer});
papers on a topic having associated data catalogs with particular 
spectral properties (\emph{exoplanets and vizier:infrared}); 
or any combination
of the two (\emph{exoplanets and (bibgroup:Spitzer or vizier:infrared)}).

Of course there are also direct benefits of this integration to the 
archives and projects providing their data to ADS.  
First of all, by having a bibliography
integrated in the ADS database, it becomes trivial to retrieve metrics associated
with it and thus evaluate the scientific impact of its datasets
as well as using bibliographic-based analytics to gain insights on how
the data is used in current research.
However, the primary reason for this activity is the scientific impact 
gained by having publications linked to data products.
Studies by \cite{2012ASPC..461..763H} and \cite{2015arXiv151102512D} have
shown a ``data sharing advantage'' for papers which have links to data products,
resulting in higher citation counts.
Similarly, data re-use increases upon the publication of papers studying
them \citep{2006ASPC..351...93W}, leading to an increase in archival 
research \citep{2009astro2010P..64W}. In other words, well linked-data is
more heavily used, and well-linked publications are more heavily cited, 
a win-win scenario, and the primary reason to have
well-described, well-curated data products indexed in ADS.

\acknowledgements This work has been supported by the NASA Astrophysics Data System project, funded by NASA grant NNX12AG54G.

\bibliography{O4-4}  

\end{document}